\documentclass[11pt,a4paper]{article}
\pdfoutput=1

\usepackage{jheppub}
\usepackage{latexsym}
\usepackage{multirow}
\usepackage{color}
\usepackage[usenames,dvipsnames,table]{xcolor}
\usepackage{graphicx}
\usepackage{subcaption}
\usepackage{epsfig} 
\usepackage{epsf}   
\usepackage{dcolumn}
\usepackage{bm}
\usepackage{dcolumn}
\usepackage{textcomp}
\usepackage{float}
\usepackage{hypcap}
\usepackage[]{hyperref}
\usepackage{makecell}
\usepackage{color}
\usepackage{pifont}
\usepackage{appendix}
\usepackage{amsmath}
\usepackage{mathrsfs}
\usepackage{multirow,bigdelim}  
\usepackage{lineno}
\usepackage[normalem]{ulem}
\graphicspath{{fig/}}
\usepackage{physics}
\usepackage{simpler-wick}
\usepackage{slashed}
\usepackage{comment}

\def\be{\begin{equation}}
\def\ee{\end{equation}}
\definecolor{maroon}{rgb}{0.5, 0.0, 0.0}	
\definecolor{arsenic}{rgb}{0.23, 0.27, 0.29}

\usepackage{amssymb}
\hypersetup{
	bookmarks=true,         
	unicode=false,          
	pdftoolbar=true,        
	pdfmenubar=true,        
	pdffitwindow=true,   
	pdfstartview={FitH},    
	pdfsubject={dark NSI},   
	pdfnewwindow=true,      
	pdfcreator={RevTeX},
	colorlinks=true,     
	linkcolor=Blue,         
	citecolor=Blue,        
	filecolor=Blue,     
	urlcolor=Blue,           
}

\usepackage{tikz,xcolor,hyperref}

\definecolor{lime}{HTML}{A6CE39}
\DeclareRobustCommand{\orcidicon}{\hspace{-4pt}
	\begin{tikzpicture}
		\draw[lime, fill=lime] (0,0) 
		circle [radius=0.16] 
		node[white] {\hspace{0.1mm}{\fontfamily{qag}\selectfont \tiny ID}};
		\draw[white, fill=white] (-0.07,0.1) 
		circle [radius=0.01];
	\end{tikzpicture}
	\hspace{-3.2mm}
}

\foreach \x in {A, ..., Z}{\expandafter\xdef\csname 
	orcid\x\endcsname{\noexpand\href{https://orcid.org/\csname orcidauthor\x\endcsname}
		{\noexpand\orcidicon}}
}

\preprint{}

\title{Resolving Lorentz-Violating New Physics at ESSnuSB Using High-Statistics Complementarity with T2HK}

\author[a, b]{Himanshu Bora\orcidA{},} 
\author[a]{Debajyoti Dutta\orcidB{},}
\author[a]{Monjowara Khatun\orcidD{},}
\author[c]{Abinash Medhi\orcidC{},}

\affiliation[a]{Department of Physics, Bhattadev University, Bajali, Pathsala, 781325, India}
\affiliation[b]{Department of Physics, Kamrup College, Chamata, Nalbari, Assam - 781306, India}
\affiliation[c]{Department of Physics, G.L. Choudhury College, Barpeta Road, Assam 781315, India}

\emailAdd{hb@kamrupcollege.ac.in}
\emailAdd{phy.debajyoti@bhattadevuniversity.ac.in}
\emailAdd{abinashmedhi0@gmail.com}
\emailAdd{monjowarak@gmail.com}

\date{\today}

\abstract{
A primary objective for next-generation long-baseline neutrino facilities is the search for Planck-scale Lorentz Invariance Violation (LIV). In this work, we explore the capabilities of the proposed ESSnuSB and T2HK experiments to constrain isotropic, CPT-violating LIV parameters ($a_{\alpha\beta}$). The modifications to oscillation probabilities induced by these LIV parameters can introduce parameter degeneracies with the atmospheric mixing angle $\theta_{23}$ and the Dirac CP-violating phase $\delta_{CP}$, which can potentially result in incorrect determination of the said standard oscillation parameters if we do not account for LIV effects. Through detailed GLoBES simulations, we find that while the second-oscillation-maximum configuration of ESSnuSB yields good constraints on the exact phase of $\delta_{CP}$, its intrinsic neutrino-antineutrino statistical asymmetry persistently leads to wrong octant fake solutions for $\theta_{23}$. By combining either the 360 km or the 540 km ESSnuSB configuration with the complementary, high-statistics first oscillation maximum measurements from T2HK's 295 km baseline, we show that the LIV-induced degeneracies are strongly reduced for most LIV parameters. Our analysis reflects how complementarity between ESSnuSB and T2HK provides an effective, matter-independent framework to break LIV-induced degeneracies and establish bounds on Planck-scale LIV physics.
}


\begin{document}
 \maketitle

\section{Introduction}\label{sec:intro}

The observation of neutrino oscillations definitively established that neutrinos possess non-zero mass, prompting the extension of the Standard Model (SM) into the three-flavor mixing framework \cite{Super-Kamiokande:1998kpq, SNO:2002tuh, KamLAND:2002uet}. Over the course of several decades, a vast array of experimental data \cite{Super-Kamiokande:2001ljr,SNO:2001kpb, SNO:2002tuh2,Super-Kamiokande:1999vcj,Super-Kamiokande:2000bnn,CHOOZ:1997zlf,DayaBay:2012fng,DoubleChooz:2012eiq,RENO:2012mkc,KamLAND:2004mhv,KamLAND:2002uet,KamLAND:2008dgz,OPERA:2018nar,T2K:2011qtm, T2K:2013ppw, NOvA:2019cyt} put stricter constraints on the neutrino oscillation parameters in this framework. Today, a whole generation of long-baseline experimental facilities including DUNE \cite{DUNE:2015lol}, T2HK \cite{Abe:2018uyc}, T2HKK \cite{Abe:2016ero}, and ESSnuSB \cite{ESSnuSB:2023design} aims at resolving the remaining critical unknowns of this paradigm by determining the neutrino mass ordering (normal or inverted), precisely measuring the leptonic CP-violating phase ($\delta_{CP}$), and uniquely resolving the octant of the atmospheric mixing angle ($\theta_{23}$). Beyond standard oscillations, the massive baselines of these experiments offer unprecedented sensitivity to Planck-scale physics, such as Lorentz Invariance Violation (LIV).

The CPT theorem dictates that any interacting quantum field theory violating CPT symmetry must necessarily also break Lorentz invariance \cite{PhysRevLett.89.231602}. Because neutrinos can traverse large distances with minimal interactions, their small Planck-suppressed LIV phase modifications can accumulate to observable levels, positioning long-baseline (LBL) experiments as exceptionally sensitive probes of Lorentz symmetry \cite{Barenboim:2018ctx, Agarwalla2020, Majhi:2019tfi, Fiza:2022xfw}. Neutrino experiments including Super-Kamiokande \cite{Super-Kamiokande:2014exs}, IceCube \cite{IceCube:2017qyp}, T2K \cite{T2K:2017ega}, MINOS \cite{MINOS:2008fnv, MINOS:2010kat}, and Double Chooz \cite{LIV_Double_Chooz_2012}, along with a broad array of general experimental searches \cite{RevModPhys.83.11} have already established stringent LIV limits.

In neutrino phenomenology, LIV effects have been extensively studied across sectors like the accelerator sector\cite{Barenboim:2018ctx, Majhi:2019tfi, Agarwalla2020, Fiza:2022xfw, Raikwal:2023lzk}, short-baseline reactor sector \cite{Abrahao:2015rba}, atmospheric sector \cite{Super-Kamiokande:2014exs, Sahoo:2022liv, IceCube:2010fyu}, solar sector \cite{SNO:2018pvg, PhysRevD.92.073003}, and astrophysical sector \cite{PhysRevD.99.043015, PhysRevD.99.043013, PhysRevD.99.123018, Cordero:2024LIVAxionESS}. In LBL contexts, LIV significantly alters sensitivities to standard oscillation parameters. For example, LIV can reduce $\theta_{23}$ octant sensitivity at DUNE, though the simultaneous presence of $a_{e\mu}$ and $a_{e\tau}$ can mutually nullify this effect \cite{Agarwalla2020}. Similarly, LIV can ease the tension between T2K and NOvA data, albeit by degrading overall standard parameter sensitivities \cite{Rahaman:2021leu}. Crucially, combining complementary experiments provides a powerful mechanism to break these LIV-induced degeneracies. This synergy has been demonstrated between NOvA and T2K \cite{Majhi:2019tfi}, DUNE and P2O \cite{Fiza:2022xfw}, DUNE and ESSnuSB \cite{Delgadillo:2025sme, Bora:2025LIV}, and through joint analyses of beam (DUNE, T2HK) and atmospheric (ICAL) data \cite{Raikwal:2023lzk}. Furthermore, exploring alternative setups, such as a High-Energy DUNE flux \cite{Giarnetti:2024universe}, highlights how harder spectra offer superior sensitivity to CPT-even ($c_{\alpha\beta}$) coefficients compared to standard beams optimized for the CPT-odd ($a_{\alpha\beta}$) parameters.

A major challenge in identifying these LIV signatures is parameter degeneracy. The non-standard phase effects induced by the $a_{\alpha\beta}$ parameters can significantly distort standard oscillation probabilities, introducing severe entanglements with the standard atmospheric mixing angle $\theta_{23}$ and the Dirac CP-violating phase $\delta_{CP}$. In a recent study \cite{Bora:2025LIV}, we demonstrated that although combining the two proposed baselines of the European Spallation Source neutrino Super Beam (ESSnuSB)—operating at 360 km and 540 km—improves overall sensitivity to LIV, it does not fully remove these degeneracies. In contrast, we established that a joint ESSnuSB+DUNE analysis, benefiting from the synergy between the second oscillation maximum sensitivity at ESSnuSB and the matter-enhanced sensitivity at DUNE at the first oscillation maximum, can successfully resolve all these degeneracies and yield significantly stronger constraints on all the LIV parameters.

Building upon that framework, this work investigates an alternative, highly promising complementary LBL configuration. DUNE relies heavily on profound matter effects across a massive 1300 km baseline. HHowever, it is crucial to understand if the resolution of these degeneracies is strictly dependent on strong matter effects or if exceptional statistical
precision at the first oscillation maximum can achieve a similar outcome. To answer this, we explore the capabilities of the proposed Hyper-Kamiokande (T2HK) experiment in this study. T2HK utilizes a 295 km baseline with a highly intense beam peaking at approximately 0.6 GeV, offering high statistics at the first oscillation maximum with comparatively smaller matter effects.

In this paper, we investigate whether incorporating the complementary, high-statistics first oscillation maximum data from T2HK also effectively breaks the parameter degeneracies observed at ESSnuSB. We demonstrate that the synergetic combination of ESSnuSB's second oscillation maximum focus and T2HK's precise first oscillation maximum measurements indeed provides a powerful, matter-independent framework to constrain Planck-scale LIV physics for most $a_{\alpha\beta}$ LIV parameters. We also compare this to the findings of our recent study \cite{Bora:2025LIV}, where we investigated the synergetic performance of ESSnuSB and DUNE.

The remainder of this work is organized as follows. In Section \ref{sec:formalism}, we briefly review the relevant Standard-Model Extension (SME) framework. The simulation configurations for T2HK and ESSnuSB are defined in Section \ref{sec:simulation}. In Section \ref{sec:sensitivity_analysis}, we present our phenomenological findings of the event rate analysis, and the sensitivity contours for the combined setup. Concluding remarks are provided in Section \ref{sec:conclusion}.

\section{Theoretical Formalism}\label{sec:formalism}

We adopt the same framework as described in our previous study \cite{Bora:2025LIV}, which we shall describe here briefly. To systematically incorporate potential violations of exact Lorentz and CPT symmetries into the Standard Model, the Standard-Model Extension (SME) introduces coordinate-independent background tensor fields \cite{Kostelecky:2003cr, Colladay:1998fq, Barenboim:2018ctx, Majhi:2019tfi, Fiza:2022xfw}. Considering only renormalizable operators (mass dimension $\le 4$ ), the modified Lagrangian density governing a family of Dirac fermions indexed by flavour indices $\alpha$ and $\beta$ is expressed as:

\begin{equation}    
\mathcal{L}_{LIV}=-\frac{1}{2}[a_{\alpha\beta}^{\mu}\overline{\psi}_{\alpha}\gamma_{\mu}\psi_{\beta}+b_{\alpha\beta}^{\mu}\overline{\psi}_{\alpha}\gamma_{5}\gamma_{\mu}\psi_{\beta}-ic_{\alpha\beta}^{\mu\nu}\overline{\psi}_{\alpha}\gamma_{\mu}\partial_{\nu}\psi_{\beta}-id_{\alpha\beta}^{\mu\nu}\overline{\psi}_{\alpha}\gamma_{5}\gamma_{\mu}\partial_{\nu}\psi_{\beta}]+h.c.
\end{equation}

In this expression, the fermion field of flavor $\alpha$ is denoted by $\psi_\alpha$. The violation of Lorentz symmetry is parameterized by the constant background fields given by the coefficients $a_{\alpha\beta}^\mu$, $b_{\alpha\beta}^\mu$, $c_{\alpha\beta}^{\mu\nu}$, and $d_{\alpha\beta}^{\mu\nu}$. Within this set,  $a_{\alpha\beta}^\mu$ and $b_{\alpha\beta}^\mu$ introduce CPT violation, while the derivative couplings governed by  $c_{\alpha\beta}^{\mu\nu}$ and $d_{\alpha\beta}^{\mu\nu}$  strictly preserve CPT symmetry \cite{Kostelecky:2003cr}.

Because standard weak interactions exclusively involve left-handed neutrinos, the physically observable LIV phenomena in the neutrino sector depend on specific linear combinations of these SME parameters \cite{Kostelecky:2003cr, Majhi:2019tfi, Agarwalla2020, Barenboim:2018ctx, Fiza:2022xfw}:
\begin{align}
(a_L)_{\alpha\beta}^\mu &= (a+b)_{\alpha\beta}^\mu \quad (\text{CPT-violating}) \\
(c_L)_{\alpha\beta}^{\mu\nu} &= (c+d)_{\alpha\beta}^{\mu\nu} \quad (\text{CPT-even})
\end{align}

Operating in flavor space, these combinations behave as Hermitian matrices. To align with standard phenomenological analyses, we restrict our study to the isotropic approximation. This assumes rotational invariance and isolates only the time-like components ($\mu=\nu=0$) \cite{Kostelecky:2003cr, Majhi:2019tfi}. Consequently, the notation simplifies to $(a_L)_{\alpha\beta}^0 \to a_{\alpha\beta}$ and $(c_L)_{\alpha\beta}^{00} \to c_{\alpha\beta}$ \cite{Majhi:2019tfi, Agarwalla2020}, defined within the conventional Sun-centered celestial equatorial frame. Our present analysis restricts its scope exclusively to the isotropic, CPT-odd coefficients .

The propagation dynamics of the three neutrino flavor states  through a matter medium are dictated by the total effective Hamiltonian:
\begin{equation}
H_{eff}=H_{vac}+H_{mat}+H_{LIV}
\label{eq:HeffTotal}
\end{equation}
This Hamiltonian is constructed from three distinct contributions:
\begin{itemize}
\item The standard kinematic vacuum evolution, $H_{vac}$ , defined as:
\begin{equation}
H_{vac}=\frac{1}{2E}U \text{diag}(0, \Delta m_{21}^{2}, \Delta m_{31}^{2}) U^{\dagger}
\label{eq:Hvac}
\end{equation}
where  U is the PMNS matrix, E is neutrino energy, and $\Delta m_{ij}^2 = m_i^2 - m_j^2$ .

\item The standard MSW potential induced by matter interactions, $H_{mat}$:
\begin{equation}
H_{mat}=\sqrt{2}G_{F}N_{e} \text{diag}(1, 0, 0)
\label{eq:Hmat}
\end{equation}
with $G_F$ representing the Fermi coupling constant and $N_e$ denoting the background electron number density.

\item  For the sake of completeness, we write the combined Hamiltonian covering both CPT-odd and CPT-even isotropic components as \cite{Kostelecky:2003cr}:
\begin{equation}
\begin{pmatrix} a_{ee} & a_{e\mu} & a_{e\tau} \\ a_{e\mu}^* & a_{\mu\mu} & a_{\mu\tau} \\ a_{e\tau}^* & a_{\mu\tau}^* & a_{\tau\tau} \end{pmatrix} - \frac{4}{3}E \begin{pmatrix} c_{ee} & c_{e\mu} & c_{e\tau} \\ c_{e\mu}^* & c_{\mu\mu} & c_{\mu\tau} \\ c_{e\tau}^* & c_{\mu\tau}^* & c_{\tau\tau} \end{pmatrix}
\end{equation}
Notably, the CPT-even part ($c_{\alpha\beta}$) scales linearly with neutrino energy $E$, whereas the CPT-odd part ($a_{\alpha\beta}$) remains entirely energy-independent \cite{Kostelecky:2003cr, Barenboim:2018ctx}. This lack of energy dependence allows the $a_{\alpha\beta}$ parameters to act essentially as static perturbations to the standard matter potential. Because we solely investigate the CPT-violating signatures in this work, the relevant new-physics Hamiltonian simplifies to:
\begin{equation}
H_{LIV} = \begin{pmatrix} a_{ee} & a_{e\mu} & a_{e\tau} \\ a_{e\mu}^{*} & a_{\mu\mu} & a_{\mu\tau} \\ a_{e\tau}^{*} & a_{\mu\tau}^{*} & a_{\tau\tau} \end{pmatrix}
\label{eq:HLIVa}
\end{equation}
The matrix Hermiticity guarantees real diagonal entries ($a_{ee}, a_{\mu\mu}, a_{\tau\tau}$) and complex off-diagonal entries such that $a_{\alpha\beta} = a_{\beta\alpha}^* = |a_{\alpha\beta}| e^{i\phi_{\alpha\beta}}$ (where $\alpha \neq \beta$). These inherent complex phases ($\phi_{e\mu}, \phi_{e\tau}, \phi_{\mu\tau}$) introduces new CP violation effects in addition to the standard PMNS phase $\delta_{CP}$.

\end{itemize}

A defining feature of the CPT-odd  operators is their transformation behavior under CP and CPT inversions. In this antineutrino framework, the LIV contribution undergoes both a sign inversion and complex conjugation. Consequently, the propagation of antineutrinos is governed by a modified effective Hamiltonian:
\begin{equation}
H_{eff}(\overline{\nu}) = H_{vac} - H_{mat}^{*} - H_{LIV}^{*}
\label{eq:HeffAntiNu}
\end{equation}

We analyze both $\nu$ and $\bar{\nu}$ oscillation channels since this difference between the neutrino and antineutrino Hamiltonians (in both $H_{mat}$ and $H_{LIV}$) is fundamental to probing CPT violation

\section{Experiment and Simulation Details}\label{sec:simulation}

To investigate the degeneracies that may arise due to isotropic, CPT-violating LIV parameters ($a_{\alpha\beta}$) at ESSnuSB and T2HK experiments, we simulate the experimental setups using the General Long Baseline Experiment Simulator (GLoBES) software package \cite{Huber:2004ka, HUBER2007439}. 

\paragraph{ESSnuSB:} The European Spallation Source neutrino Super Beam (ESSnuSB) is an upcoming LBL experiment designed to deliver a highly intense neutrino beam produced at the European Spallation Source (ESS) at Lund, Sweden driven by a proposed 5 MW (upgradable to 10 MW) proton linac \cite{ESSnuSB:2023design}. We simulate two potential far-detector sites housing MEMPHYS-like Water Cherenkov detectors (538 kton fiducial mass): Zinkgruvan at a baseline of 360 km, and Garpenberg at 540 km. The proposed setup will produce a neutrino beam spectrum between 0.2 GeV and 0.6 GeV, peaking right around 0.35 GeV. At the 540 km Garpenberg baseline, the $L/E$ ratio places the 0.35 GeV peak at the second oscillation maximum for $\nu_\mu \rightarrow \nu_e$ appearance. On the other hand, for the 360 km Zinkgruvan baseline, the beam spectrum allows the experiment to sample the second oscillation maximum via the low-energy tail ($\sim$0.25 GeV) and the first oscillation maximum via the high-energy tail ($\sim$0.6 GeV). We refer the readers to our previous work \cite{Bora:2025LIV} for plots of the event spectrum for both these baselines. We assume a 10-year runtime, divided equally between neutrino and antineutrino operation modes.

\paragraph{T2HK:} The Tokai to Hyper-Kamiokande (T2HK) experiment is a next-generation LBL neutrino observatory currently under construction in Japan \cite{Abe:2018uyc}. It will utilize an upgraded, high-intensity narrowband off-axis neutrino beam generated by the Japan Proton Accelerator Research Complex (J-PARC), planned to operate at a beam power of 1.3 MW \cite{Abe:2018uyc}. The beam spectrum is sharply peaked at an energy of $E_\nu \sim 0.6$ GeV, optimally tuned to the first oscillation maximum for high statistics. The far detector is a massive next-generation Water Cherenkov detector with a fiducial mass of 187 kton, situated 295 km away from the neutrino source in the Kamioka mine \cite{Abe:2018uyc}. In contrast to DUNE, another LBL experiment, its relatively short baseline (295 km) results in comparatively small matter effects. For T2HK, we assume a total exposure of 10 years with a neutrino-to-antineutrino runtime ratio of 1:3 (2.5 years in neutrino mode and 7.5 years in antineutrino mode).

In this work, we report the individual capabilities of the ESSnuSB's 360 km baseline and 540 km baseline configurations as well as that of T2HK's 295 km baseline. Our analysis focuses next on the synergetic performances of two distinct configurations - 
\begin{itemize}
    \item ESSnuSB (540 km baseline) and T2HK (295 km baseline),
    \item ESSnuSB (360 km baseline) and T2HK (295 km baseline)
\end{itemize}

Since CPT-violating $a_{\alpha\beta}$ parameters enter the antineutrino effective Hamiltonian with a reversed sign and complex conjugation (Eq \ref{eq:HeffAntiNu}), comparing $\nu$ and $\bar{\nu}$ transitions is necessary for investigating LIV effects. Both the simulated experiments feature dedicated neutrino and antineutrino runs. The key input parameters for the GLoBES simulations used in our analysis for the ESSnuSB and T2HK experiments are summarized in Table \ref{tab:uncertainity}. We outline the physical configurations—including baselines, $L/E$ ratios, detector masses, and runtime split between neutrino and antineutrino modes. We also mention the assumed systematic normalization uncertainties for signal and background events for neutrino channels (un-bracketed) and antineutrino channels (bracketed).

\begin{table}[H]
    \centering
    \renewcommand{\arraystretch}{1.2}
    \begin{tabular}{|c|c|c|c|}
        \hline
        \multirow{2}{*}{Experiment details} &\multirow{2}{*}{Channels} & \multicolumn{2}{|c|}{Normalization uncertainty} \\ \cline{3-4}
        & & Signal & Background \\
        \hline \hline
        \textbf{ESSnuSB}, Baseline: 360 \& 540 km    &  & & \\
         L/E = 1200 \& 1534 km/GeV &  $\nu_e (\bar \nu_e)$ appearance  & 5\% (5\%) & 5\% (5\%) \\ 
         Fiducial mass = 538 kt (WC)& $\nu_\mu (\bar \nu_\mu)$ disappearance  & 5\% (5\%) & 5\% (5\%)  \\ 
         Runtime : 5 years $\nu$ + 5 years $\bar \nu$  & & & \\
        \hline
         \textbf{T2HK}, Baseline: 295 km  &  & &\\
         L/E = $\sim 492$ km/GeV &  $\nu_e (\bar \nu_e)$ appearance & 3.2\% (3.9\%) & 3.2\% (3.9\%) \\
         Fiducial mass = 187 kt (WC) & $\nu_\mu (\bar \nu_\mu)$ disappearance & 3.6\% (3.6\%) & 3.6\% (3.6\%) \\ 
         Runtime : 2.5 years $\nu$ + 7.5 years $\bar \nu$  & & & \\
         \hline
    \end{tabular}
    \caption{Experimental configurations and detector uncertainties for the simulated ESSnuSB and T2HK setups used in our GLoBES simulation.}
    \label{tab:uncertainity}
\end{table}

\textbf{Parameter Values and Marginalization:} 
Table \ref{tab:marginalization_val} summarizes the true values and marginalization ranges for the standard neutrino oscillation parameters used in our simulations. These inputs are derived from the latest NuFIT-6.0 global analysis \cite{Esteban:2024eli}, supplemented by recent JUNO constraints on the solar mixing angle $\theta_{12}$ \cite{Abusleme2025JUNO}. We conduct our primary analysis assuming a true normal mass ordering (NO). For the new physics sector, the true hypothesis assumes purely standard oscillations (setting all $a_{\alpha\beta}$ and $\phi_{\alpha\beta}$ to zero), whereas these LIV coefficients are marginalized over relevant ranges during the test hypothesis fitting. Benchmark LIV parameter magnitudes of $\mathcal{O}(10^{-23})$ GeV are used for illustration, motivated by our previous work \cite{Bora:2025LIV}). Specifically, the simulations utilize $a_{ee} = 2.4, a_{\mu\mu} = 3.0, a_{\tau\tau} = 2.0, |a_{e\mu}| = 0.7, |a_{e\tau}| = 1.2$, and $|a_{\mu\tau}| = 2.0$, all in units of $10^{-23}$ GeV. 

\begin{table*}[!h]
    \centering 
    \renewcommand{\arraystretch}{1.3}
    \begin{tabular}{|c|c|c|}
    \hline 
    \rule{0pt}{12pt} Oscillation Parameter & Best Fit Value & Details of Marginalization \tabularnewline
    \hline \hline
    \rule{0pt}{12pt} $\theta_{12}$       & $33.68^{\circ}$               &  Prior from JUNO \cite{Abusleme2025JUNO} \tabularnewline \hline 
    \rule{0pt}{12pt} $\theta_{13}$       & $8.52^{\circ}$                & within 1$\sigma$ allowed range \tabularnewline \hline 
    \rule{0pt}{12pt} $\theta_{23}$       & $48.5^{\circ}$                & within 3$\sigma$ allowed range\tabularnewline \hline 
    \rule{0pt}{12pt} $\delta_{CP}$       & $-90^{\circ}$                 & $[-180^{\circ},180^{\circ}]$\tabularnewline \hline 
    \rule{0pt}{12pt} $\Delta m_{21}^{2}$ & $7.49\times10^{-5}\rm~eV^{2}$ & within 1$\sigma$ allowed range \tabularnewline \hline
    \renewcommand{\arraystretch}{1.0}
    \rule{0pt}{12pt} $\Delta m_{31}^{2}$ &  $2.53\times10^{-3}\rm~eV^{2}$& $\left( [-2.58,-2.44] \cup [2.46,2.60] \right)\times10^{-3}\rm~eV^{2}$\tabularnewline  
    
    \hline 
    \end{tabular}\\
    \vspace{0.2cm}
    \caption{The oscillation parameter values used in our analysis, along with the corresponding marginalization details (NuFit 6.0 \cite{Esteban:2024eli}). }
    \label{tab:marginalization_val}
\end{table*} 

The total $\Delta\chi^2$ test statistic is formulated incorporating statistical fluctuations, systematic uncertainties (via the pull method), and prior constraints on known parameters as detailed in Ref.~\cite{Fiza:2022xfw}:

\begin{equation}
\begin{split}
    \Delta\chi^{2}(p^{\text{true}}) = \min_{p^{\text{test}},\eta} \Bigg[ & 2\sum_{i,j,k}^{} \left\{ N_{ijk}^{\text{test}}(p^{\text{test}};\eta) - N_{ijk}^{\text{true}}(p^{\text{true}}) + N_{ijk}^{\text{true}}(p^{\text{true}}) \ln\frac{N_{ijk}^{\text{true}}(p^{\text{true}})}{N_{ijk}^{\text{test}}(p^{\text{test}};\eta)} \right\} \\
    & + \sum_{l} \frac{(p_{l}^{\text{true}} - p_{l}^{\text{test}})^{2}}{\sigma_{p_{l}}^{2}} + \sum_{m} \frac{\eta_{m}^{2}}{\sigma_{\eta m}^{2}} \Bigg]
\end{split}
\end{equation}

Here, $N^{\text{true/test}}_{ijk}$ denotes the expected event rates across energy bins ($i$), oscillation channels ($j$), and beam polarities ($k$). The second term applies the parameter priors ($\sigma_{p_l}$), while the final term uses the pull method to incorporate systematic normalization uncertainties ($\eta_m$), as defined in Table \ref{tab:uncertainity}. The minimization is performed simultaneously over the relevant test parameters ($p^{\text{test}}$) and systematic nuisance parameters ($\eta$).

\section{Sensitivity Analysis : Resolving $\theta_{23}$ and $\delta_{CP}$ Degeneracies} \label{sec:sensitivity_analysis}

\subsection{Bi-Event Analysis} \label{sec:bievent}

\begin{figure}[htbp]
  \centering
  \begin{subfigure}[b]{0.39\textwidth}
    \centering
    \includegraphics[width=\textwidth]{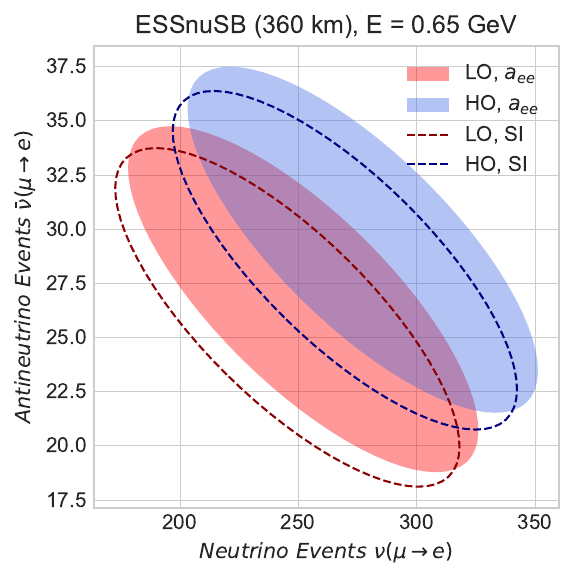}
    \label{fig:subA}
  \end{subfigure}
  \begin{subfigure}[b]{0.39\textwidth}
    \centering
    \includegraphics[width=\textwidth]{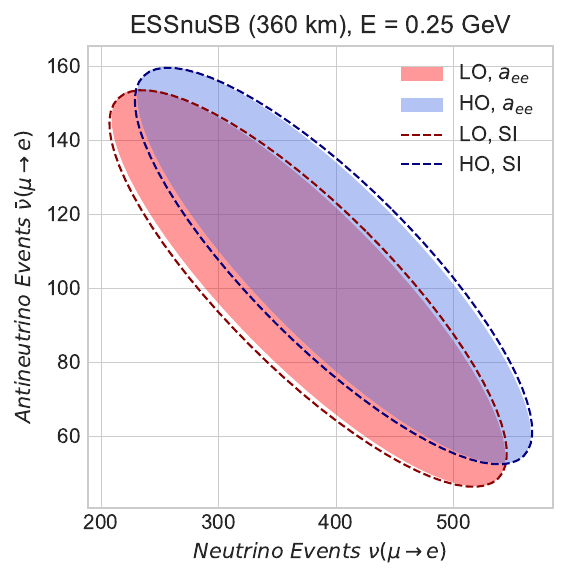}
    \label{fig:subB}
  \end{subfigure}
  \\[-4ex]

  \begin{subfigure}[b]{0.39\textwidth}
    \centering
    \includegraphics[width=\textwidth]{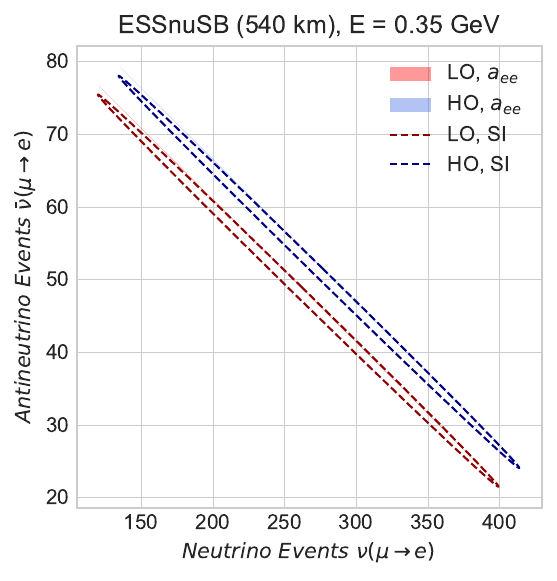}
    \label{fig:subC}
  \end{subfigure}
  \begin{subfigure}[b]{0.39\textwidth}
    \centering
    \includegraphics[width=\textwidth]{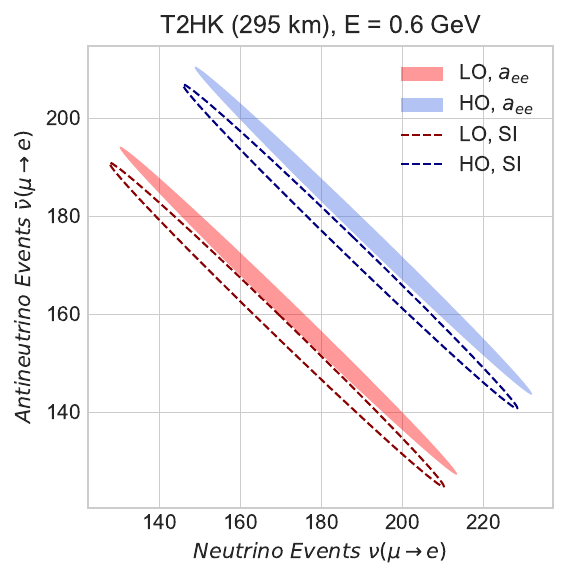}
    \label{}
  \end{subfigure}
  \\[-4ex]
  \caption{Bi-event Plot for diagonal LIV parameter $a_{ee} = 2.4 \times 10^{-23}$ GeV. The expected antineutrino appearance events ($\bar{\nu}_{\mu} \rightarrow \bar{\nu}_{e}$) are plotted against the neutrino appearance events ($\nu_{\mu} \rightarrow \nu_{e}$) at their corresponding baseline peak energies varying $\delta_{CP}$ continuously over its full range $[-\pi, \pi]$. Dashed contours represent the Standard Interaction (SI) framework ($a_{ee} = 0$) and shaded regions represent effects of LIV for Higher Octant (HO, blue) and Lower Octant (LO, red) of the atmospheric mixing angle $\theta_{23}$. }
  \label{fig:event_a11}
\end{figure}

\begin{figure}[!h]
  \centering
  \begin{subfigure}[b]{0.39\textwidth}
    \centering
    \includegraphics[width=\textwidth]{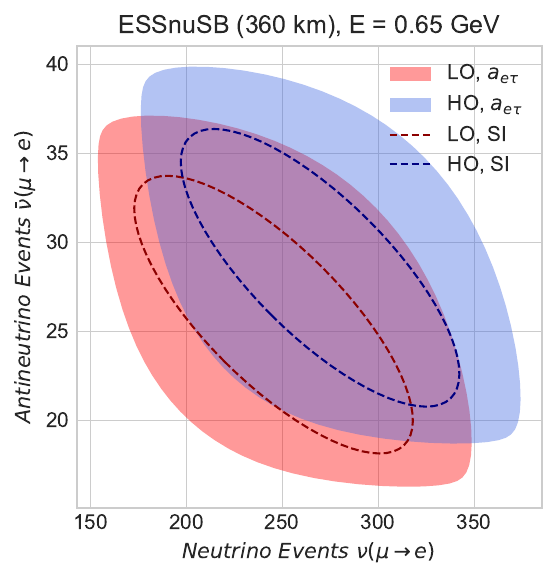}
    \label{fig:subA}
  \end{subfigure}
  \begin{subfigure}[b]{0.39\textwidth}
    \centering
    \includegraphics[width=\textwidth]{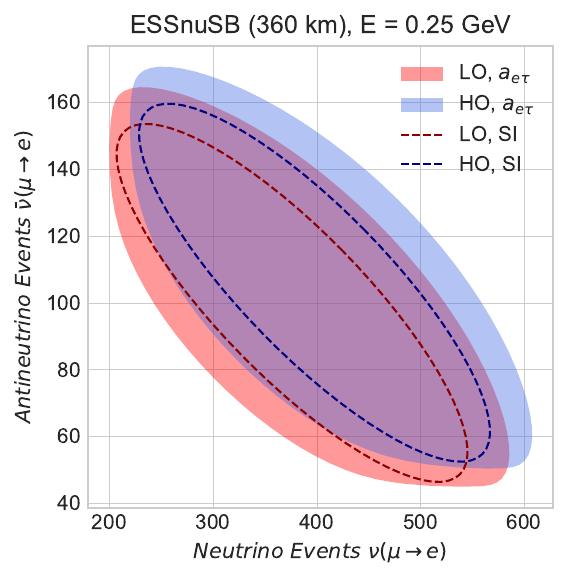}
    \label{fig:subB}
  \end{subfigure}
  \\[-4ex]

  \begin{subfigure}[b]{0.39\textwidth}
    \centering
    \includegraphics[width=\textwidth]{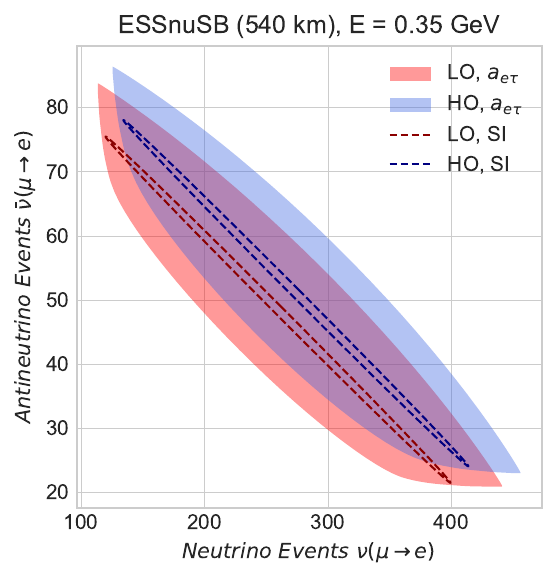}
    \label{fig:subC}
  \end{subfigure}
  \begin{subfigure}[b]{0.39\textwidth}
    \centering
    \includegraphics[width=\textwidth]{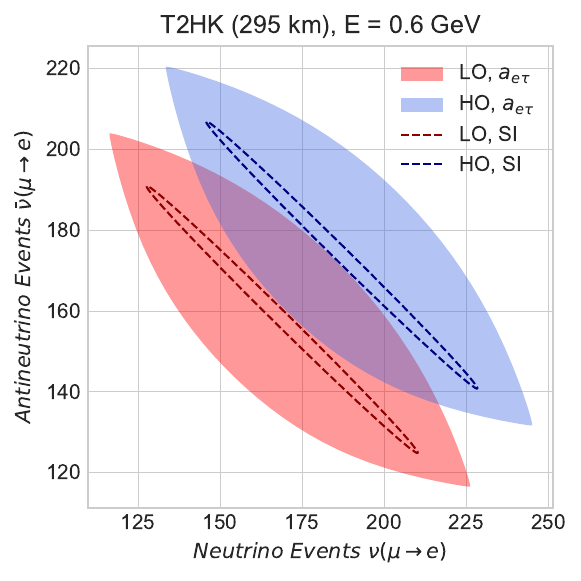}
    \label{}
  \end{subfigure}
  \\[-4ex]

  \caption{Bi-event Plot for off-diagonal LIV parameter $a_{e\tau} = 0.7 \times 10^{-23} $ GeV. The expected antineutrino appearance events ($\bar{\nu}_{\mu} \rightarrow \bar{\nu}_{e}$) are plotted against the neutrino appearance events ($\nu_{\mu} \rightarrow \nu_{e}$) at their corresponding baseline peak energies varying $\delta_{CP}$ and $\phi_{e\tau}$ continuously over their full range $[-\pi, \pi]$. Dashed contours represent the Standard Interaction (SI) framework ($a_{e\tau} = 0$) and shaded regions represent effects of LIV for Higher Octant (HO, blue) and Lower Octant (LO, red) of the atmospheric mixing angle $\theta_{23}$.}
  \label{fig:event_a13}
\end{figure}

\begin{figure}[!h]
  \centering
  \begin{subfigure}[b]{0.39\textwidth}
    \centering
    \includegraphics[width=\textwidth]{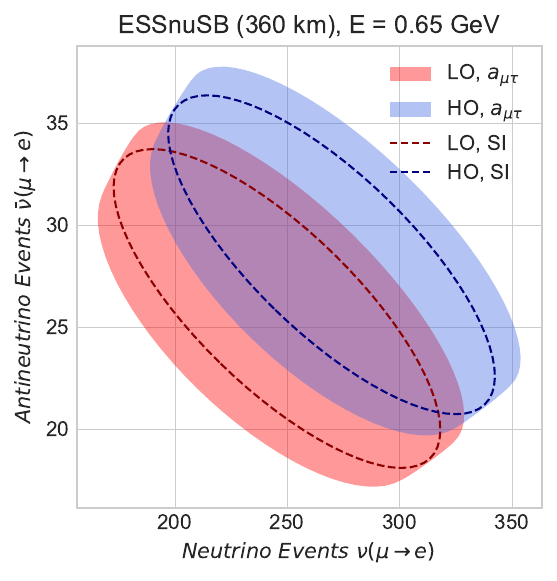}
    \label{fig:subA}
  \end{subfigure}
  \begin{subfigure}[b]{0.39\textwidth}
    \centering
    \includegraphics[width=\textwidth]{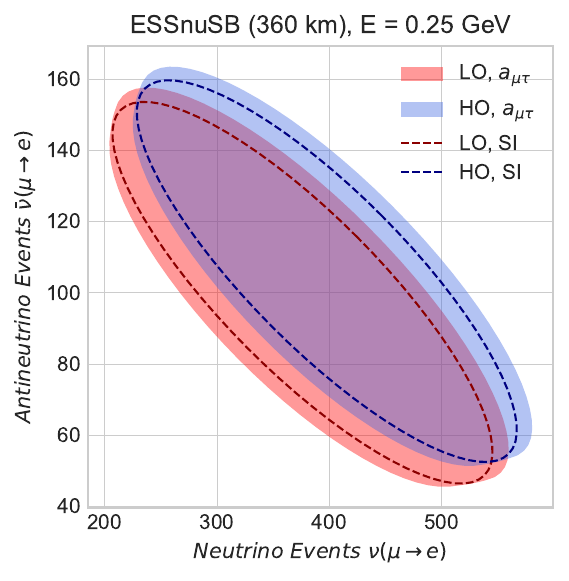}
    \label{fig:subB}
  \end{subfigure}
  \\[-4ex]

  \begin{subfigure}[b]{0.39\textwidth}
    \centering
     \includegraphics[width=\textwidth]{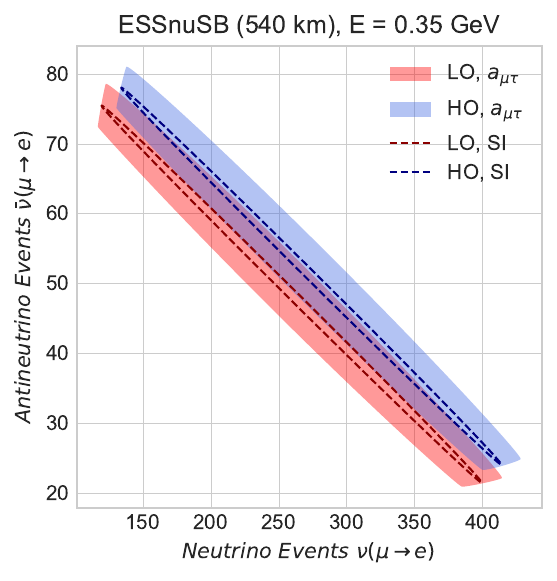}
    \label{fig:subC}
  \end{subfigure}
  \begin{subfigure}[b]{0.39\textwidth}
    \centering
    \includegraphics[width=\textwidth]{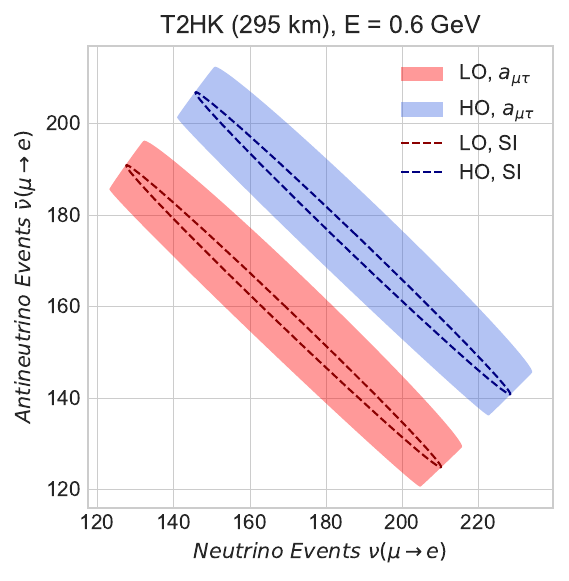}
    \label{}
  \end{subfigure}
  \\[-4ex]

  \caption{Bi-event Plot for off-diagonal LIV parameter $a_{\mu\tau} = 2.0 \times 10^{-23}$ GeV. The expected antineutrino appearance events ($\bar{\nu}_{\mu} \rightarrow \bar{\nu}_{e}$) are plotted against the neutrino appearance events ($\nu_{\mu} \rightarrow \nu_{e}$) at their corresponding baseline peak energies varying $\delta_{CP}$ and $\phi_{\mu\tau}$ continuously over their full range $[-\pi, \pi]$. Dashed contours represent the Standard Interaction (SI) framework ($a_{\mu\tau} = 0$) and shaded regions represent effects of LIV for Higher Octant (HO, blue) and Lower Octant (LO, red) of the atmospheric mixing angle $\theta_{23}$.}
  \label{fig:event_a23}
\end{figure}

We perform an analysis of the neutrino and antineutrino event rates simultaneously to intuitively get an idea of the parameter degeneracies and see if the possibility of its resolution is likely. For illustration purposes, we choose to  plot the expected antineutrino appearance events ($\bar{\nu}_{\mu} \rightarrow \bar{\nu}_{e}$) against the neutrino appearance events ($\nu_{\mu} \rightarrow \nu_{e}$) at their corresponding baseline peak energies to display the bi-event plots of the different experimental configurations for the $a_{ee}$ diagonal parameter (Figure \ref{fig:event_a11}), and $a_{e\tau}$ (Figure \ref{fig:event_a13}) and $a_{\mu\tau}$ (Figure \ref{fig:event_a23}) off-diagonal parameters. We choose $E=0.65$ GeV and $0.25$ GeV for representing two slices of the ESSnuSB 360 km baseline's first and second oscillation maximums, $E=0.35$ GeV for the second oscillation maximum of the ESSnuSB 540 km baseline and $E=0.6$ GeV for the T2HK 295 km baseline's first oscillation maximum.

These parametric curves are generated by varying the Dirac CP-violating phase $\delta_{CP}$ continuously over its full range $[-\pi, \pi]$ for the diagonal parameters and varying both $\delta_{CP}$ and LIV phase $\phi_{\alpha \beta}$ continuously over their full range $[-\pi, \pi]$ for the off-diagonal parameters, which traces out closed ellipses in the bi-event space. In these plots, the dashed contours represent the Standard Interaction (SI) framework assuming exact Lorentz symmetry ($a_{\alpha\beta} = 0$). The shaded regions illustrate the event space under the test hypothesis incorporating specific, non-zero LIV parameters. To explore the octant degeneracy, the shaded regions are plotted for both the Higher Octant (HO, blue) and Lower Octant (LO, red) of the atmospheric mixing angle $\theta_{23}$.

As revealed by the axes of these bi-event plots, there is a stark difference in event rate scales between the experiments. For ESSnuSB (at both baselines), the expected antineutrino events are significantly lower than the neutrino events—often by nearly an order of magnitude, for e.g., $\sim 300$ $\nu$ events vs. $\sim 30$ $\bar{\nu}$ events at the 360 km baseline for $E = 0.65$ GeV owing to equal runtimes of 5 years for both neutrino and antineutrino channels, antineutrino events being much lower in number for an equal runtime. Since isolating CPT-violating LIV signatures depends fundamentally on measuring the difference between neutrino and antineutrino transitions, this statistical imbalance severely limits ESSnuSB's ability to rigidly constrain the parameters. We recall that the beam fluxes in neutrino and antineutrino modes, together with the chosen exposure in each mode, are designed with the runtime ratio of 1:3 ($\nu:\bar\nu$) such that the accumulated neutrino and antineutrino event statistics are much more comparable. Hence, T2HK gives us roughly balanced and high statistics across both channels (e.g., $\sim 180$ events for both $\nu$ and $\bar{\nu}$). At the second oscillation maximum (ESSnuSB 360 km and 540 km), the intrinsic CP asymmetry is amplified, resulting in SI ellipses with a different slope compared to the T2HK SI ellipses. The symmetric statistics of T2HK restricts the T2HK contours into the narrow, distinct diagonal bands observed in the plots.

For the diagonal parameter $a_{ee}$ (Figure \ref{fig:event_a11}), no new CP-violating phase is introduced as diagonal parameters are strictly real and act as static, energy-independent background potentials. We see their primary effect in the shaded regions as a shift in the standard ellipses in the bi-event plane and slight shrinkage or expansion. At the first and second oscillation maximum of the ESSnuSB 360 km baseline, the intrinsic CP asymmetry is amplified, resulting in naturally broad SI ellipses. These SI ellipses are overlapping which indicate the presence of octant degeneracy even without the presence of LIV parameters. When the static LIV perturbation is introduced, these broad ellipses shift and change shape, causing the red (LO) and blue (HO) shaded regions to overlap significantly. An observed event rate contained within this intersected area could be equally interpreted as a standard LO signal or an LIV-distorted HO signal. Notably, the SI ellipses as well as LIV ellipses at the second oscillation maximum of the ESSnuSB 540 km baseline are very narrow despite being non-overlapping. The LIV regions have been squished to an almost invisible thin line shifted slightly upwards diagonally. Although the non-overlapping ellipses representing higher and lower octants hint at well-resolved octants both in absence and presence of the LIV parameter $a_{ee}$, the extremely narrow SI ellipse and the even narrower LIV regions located very closely with each other pose a problem in inferring true octant information in actual event counting data from the experiments. Due to experimental uncertainties, it is likely that an actual data point comprising of neutrino and antineutrino events might not lie within the thin ellipses and shaded regions. We might, hence, lose certainty in telling apart the higher octant from the lower octant or LIV effects from standard interaction as all the shapes are very close to each other despite being non-overlapping for lower and higher octants. We also note that the neutrino beam peaks around the 0.35 GeV peak for the 540 km baseline and the event number asymmetry at this point is huge; the number of neutrino events is almost five times greater than that of the antineutrino events. This adds more to the uncertainty in resolving the octants for the narrow and closely located ellipses as a slight miscount in antineutrino events may significantly reduce our ability to truly tell apart the octants. On the other hand, T2HK's high statistics across both modes is symmetric (e.g., $\sim 180$ events for both $\nu$ and $\bar{\nu}$) and the SI ellipses and LIV regions are located far apart despite them being narrow. The symmetric statistics of neutrino and antineutrino modes for the distinct non-overlapping SI ellipses and LIV regions located far apart makes T2HK much more likely to resolve the octant degeneracy compared to the 540 km baseline configuration of ESSnuSB. We also expect it to provide a significant contribution to the synergetic analysis aiming to resolve the octant degeneracy for the same reasons.

Unlike their diagonal counterparts, the complex parameters $a_{e\tau}$ (Figure \ref{fig:event_a13}) and $a_{\mu\tau}$ (Figure \ref{fig:event_a23}) introduce entirely new non-standard phases ($\phi_{e\tau}$) and ($\phi_{\mu\tau}$) respectively. Since the simulation marginalizes over these new phases in the full range $[-\pi, \pi]$, this extra degree of freedom severely distorts the allowed bi-event regions even if SI ellipses remain as non-overlapping. At ESSnuSB, this interference causes significant smearing of the parameter space, leading to almost total overlap between the HO and LO regions. The ellipses merge into large degenerate blobs where standard CP and new-physics CP effects become indistinguishable. The bi-event plot for T2HK's $0.6$ GeV oscillation peak shows overlap of the octants in presence of LIV for the  $a_{e\tau}$ parameter (Figure \ref{fig:event_a13}) while the octants for the $a_{\mu\tau}$ parameter (Figure \ref{fig:event_a23}) are distinctly non-overlapping. We expect good resolution of the octant degeneracy for $a_{\mu\tau}$ but not for the $a_{e\tau}$ parameter at T2HK. 

Because ESSnuSB and T2HK operate at different $L/E$ ratios with vastly different statistics, their bi-event ellipses possess entirely different slopes and dependencies on the LIV parameters. This crude bi-event analysis at a few energy slices clearly highlights the necessity of the multi-baseline synergetic approach if we want to break the parameter degeneracies.

\subsection{ $\chi ^2$ Analysis} \label{sec:chi_2}

Potential degeneracies between LIV parameters and the standard oscillation parameters pose a critical challenge in the search for new physics. Such degeneracies can lead to incorrect physical interpretations by obscuring the true values of the parameters. We analyse the correlation between the CPT-violating LIV parameters and the atmospheric mixing angle, $\theta_{23}$ as well as the standard CP-violating phase $\delta_{CP}$, to assess whether the different experimental configurations that we consider can resolve these degeneracies. In generating the simulated ``true'' data, the true values are fixed at $\theta_{23}=48.5^\circ$ and $\delta_{CP}=-90^\circ$. However, in the test hypothesis used to obtain the allowed regions, the relevant standard oscillation parameters are marginalized according to the ranges listed in Table 2. For Fig. 4, the plotted plane is $(\theta_{23}, a_{\alpha\beta})$. Therefore, $\theta_{23}$ is scanned/plotted along the vertical axis, while $\delta_{CP}$ is marginalized over its allowed range. For Fig. 5, the plotted plane is $(\delta_{CP}, a_{\alpha\beta})$ and so $\delta_{CP}$ is scanned/plotted along the vertical axis, while $\theta_{23}$ is marginalized over its allowed range. For the off-diagonal LIV parameters, the corresponding LIV phase $\phi_{\alpha\beta}$ is also marginalized over its full range.

\begin{figure}[!h]
  \centering
  \begin{subfigure}[b]{0.48\textwidth}
    \centering
    \includegraphics[width=\textwidth]{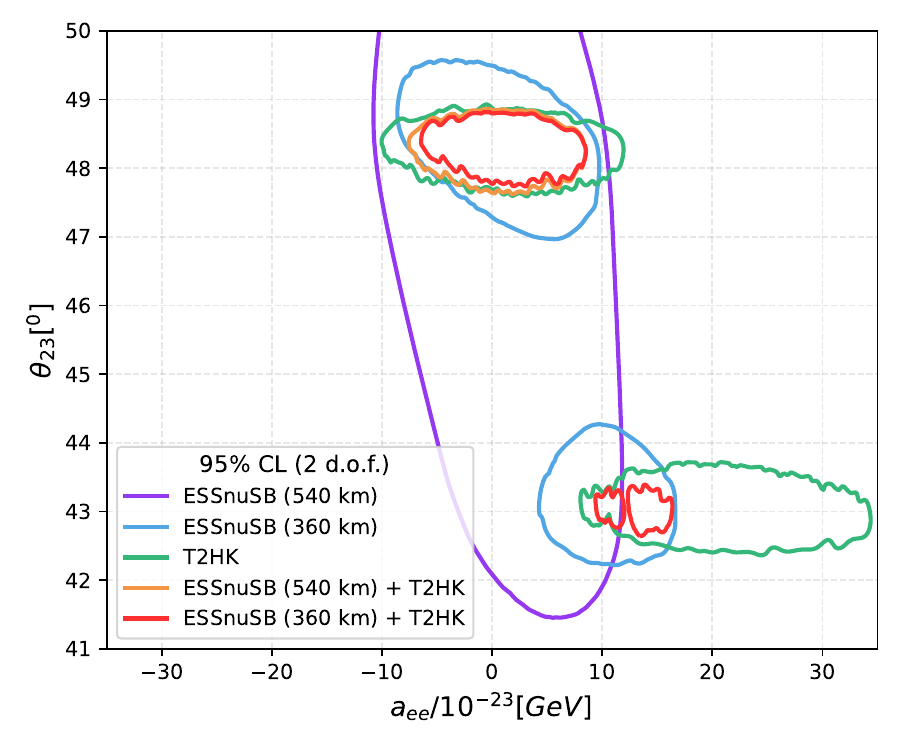}
    \label{fig:subA}
  \end{subfigure}
  \begin{subfigure}[b]{0.48\textwidth}
    \centering
    \includegraphics[width=\textwidth]{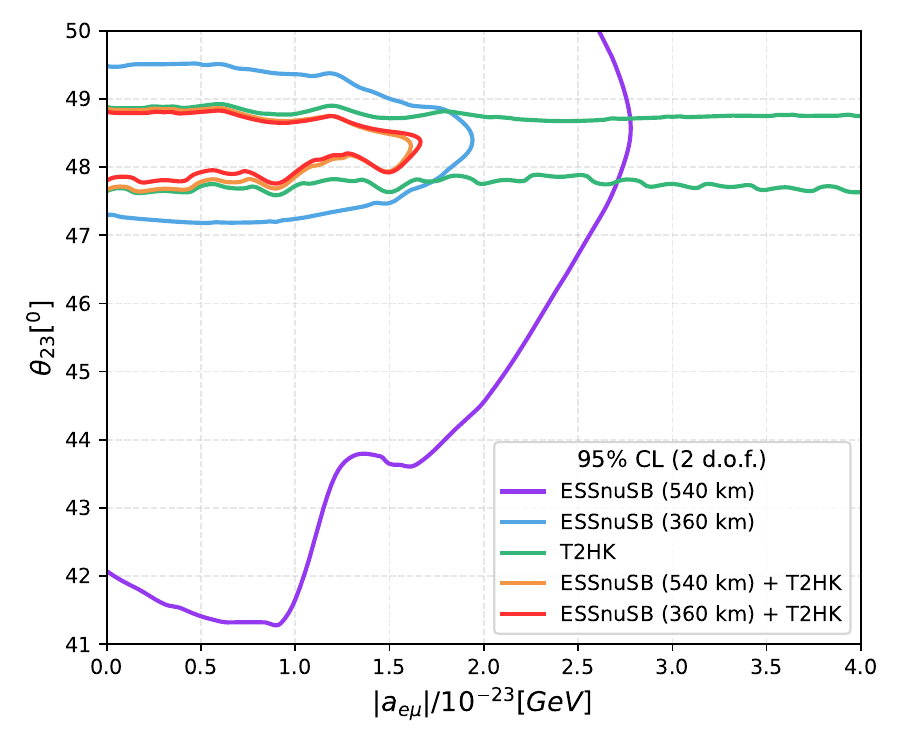}
    \label{fig:subB}
  \end{subfigure}
  \\[-4ex]

  \begin{subfigure}[b]{0.48\textwidth}
    \centering
    \includegraphics[width=\textwidth]{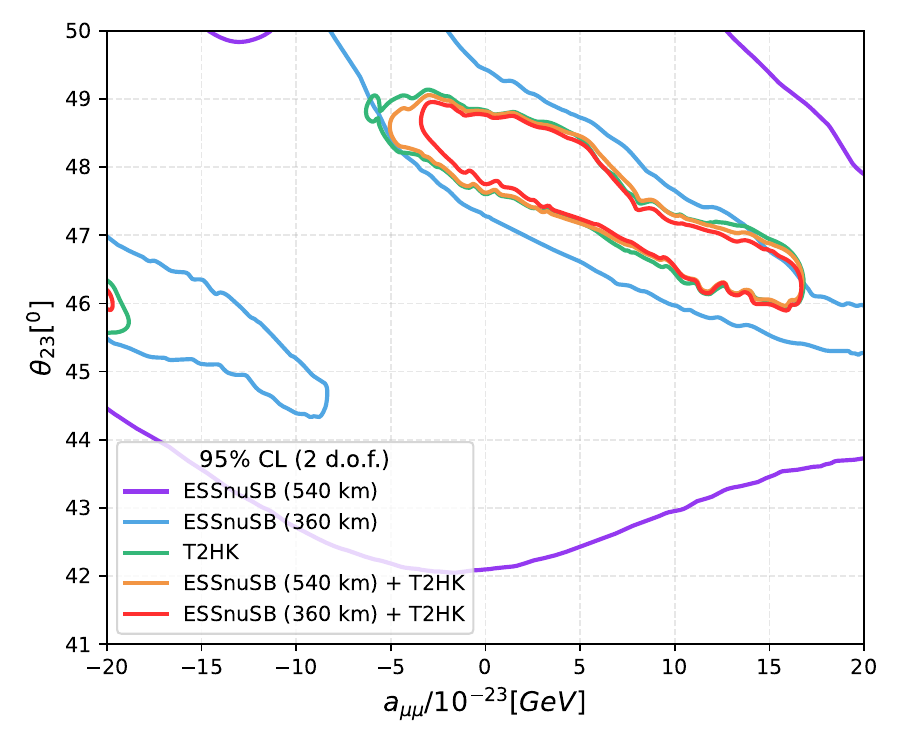}
    \label{fig:subC}
  \end{subfigure}
  \begin{subfigure}[b]{0.48\textwidth}
    \centering
    \includegraphics[width=\textwidth]{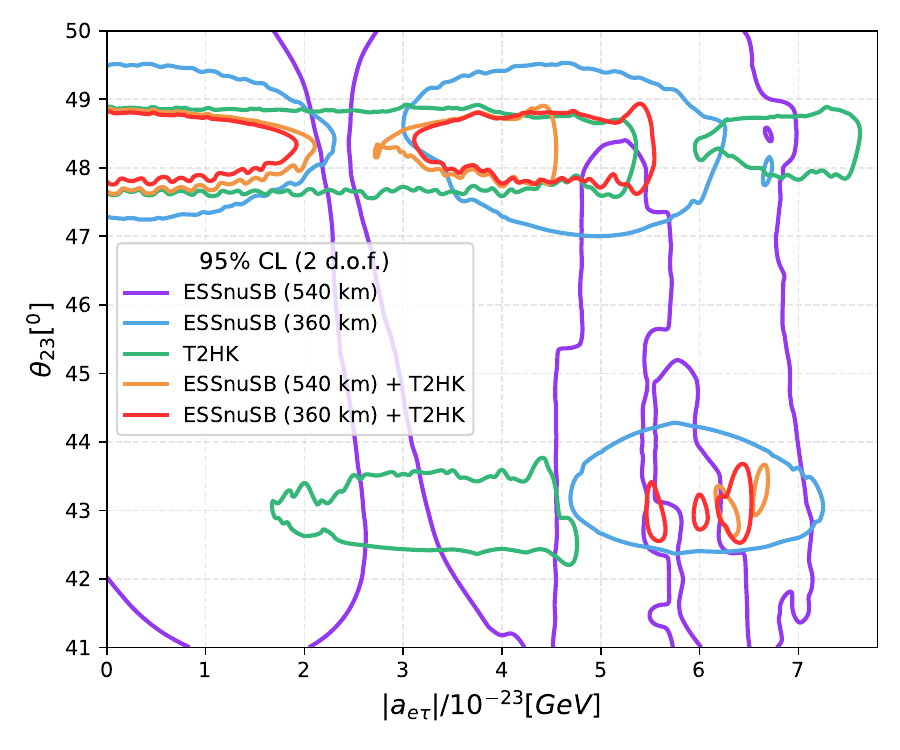}
    \label{fig:subD}
  \end{subfigure}
  \\[-4ex]

  \begin{subfigure}[b]{0.48\textwidth}
    \centering
    \includegraphics[width=\textwidth]{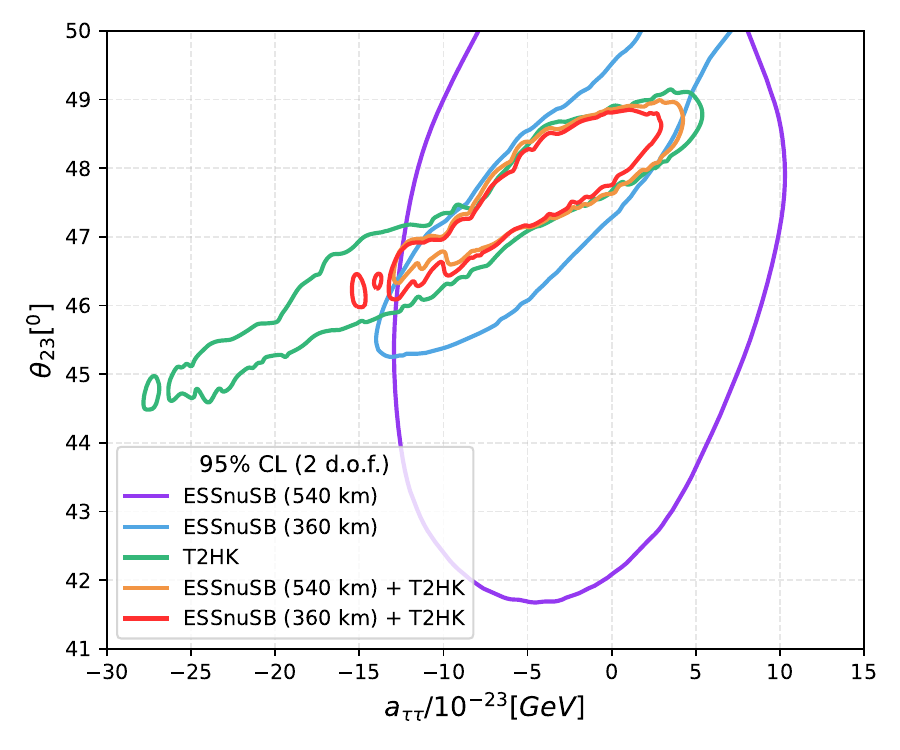}
    \label{fig:subE}
  \end{subfigure}
  \begin{subfigure}[b]{0.48\textwidth}
    \centering
    \includegraphics[width=\textwidth]{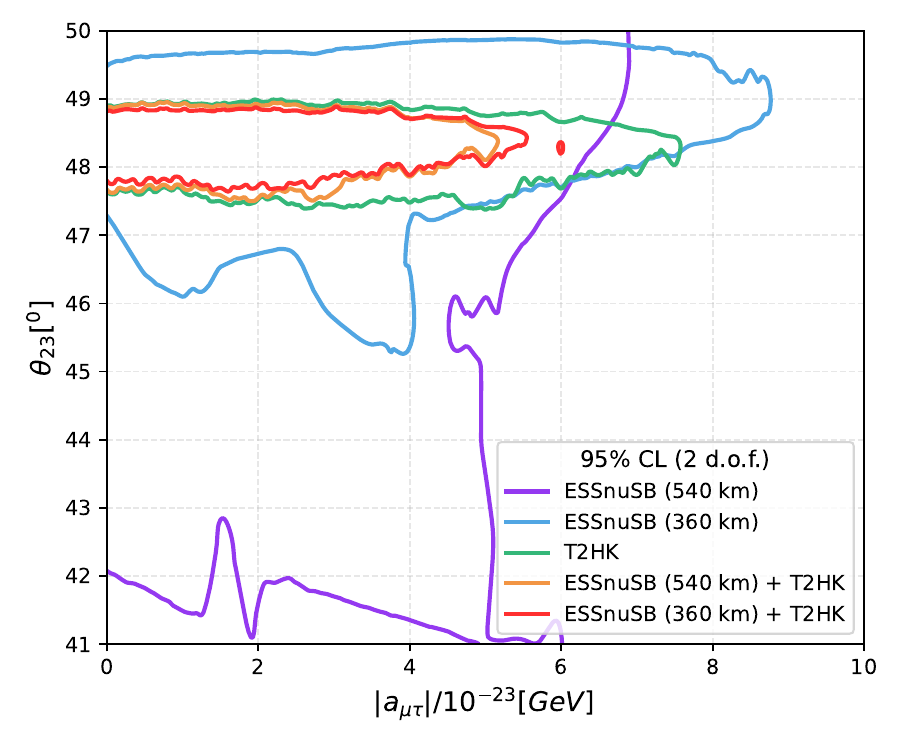}
    \label{fig:subF}
  \end{subfigure}
  \\[-4ex]

\caption{Allowed regions at 95\% C.L. (2 d.o.f.) in the plane of each LIV parameter versus $\theta_{23}$. The contours show the sensitivity for ESSnuSB for BL=540 km (purple), ESSnuSB for BL=360 km (cyan), T2HK (green), the combined analysis of ESSnuSB for BL=540 km and T2HK (orange), and the combined analysis of ESSnuSB for BL=360 km and T2HK (red). The true value of all LIV parameters is assumed to be zero.}
  \label{fig:theta23_degeneracy}
\end{figure}

\begin{figure}[!h]
  \centering
  \begin{subfigure}[b]{0.48\textwidth}
    \centering
    \includegraphics[width=\textwidth]{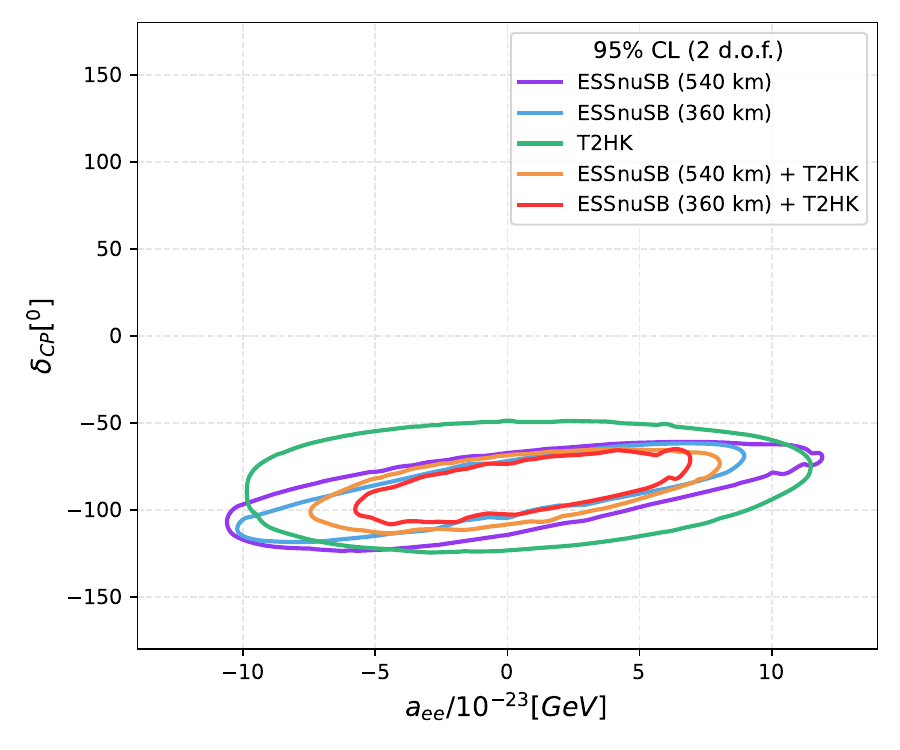}
    \label{fig:subA}
  \end{subfigure}
  \begin{subfigure}[b]{0.48\textwidth}
    \centering
    \includegraphics[width=\textwidth]{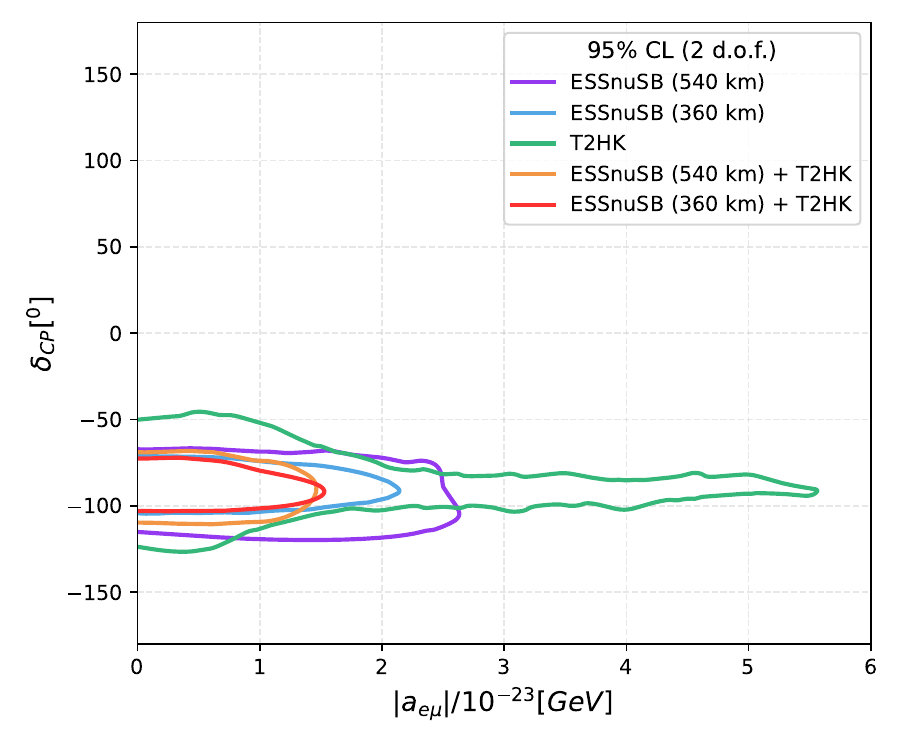}
    \label{fig:subB}
  \end{subfigure}
  \\[-4ex]

  \begin{subfigure}[b]{0.48\textwidth}
    \centering
    \includegraphics[width=\textwidth]{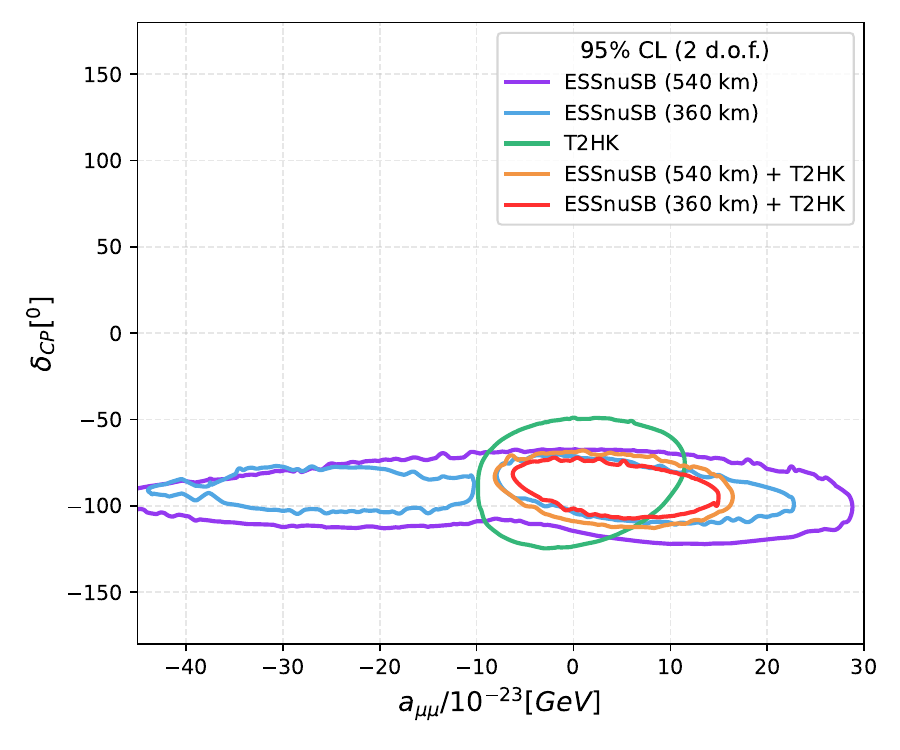}
    \label{fig:subC}
  \end{subfigure}
  \begin{subfigure}[b]{0.48\textwidth}
    \centering
    \includegraphics[width=\textwidth]{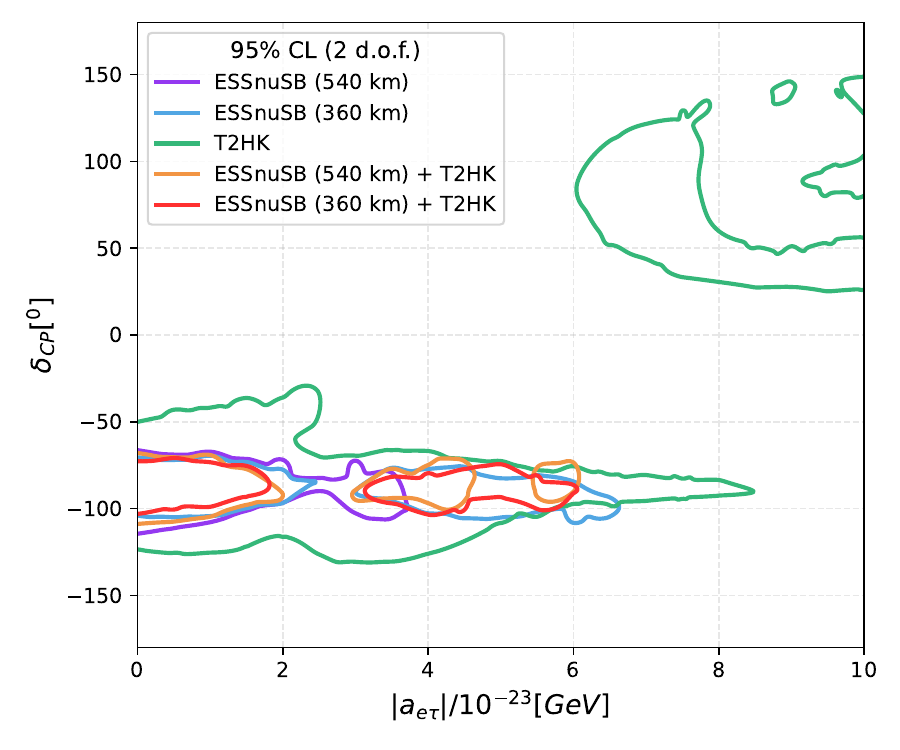}
    \label{fig:subD}
  \end{subfigure}
  \\[-4ex]

  \begin{subfigure}[b]{0.48\textwidth}
    \centering
    \includegraphics[width=\textwidth]{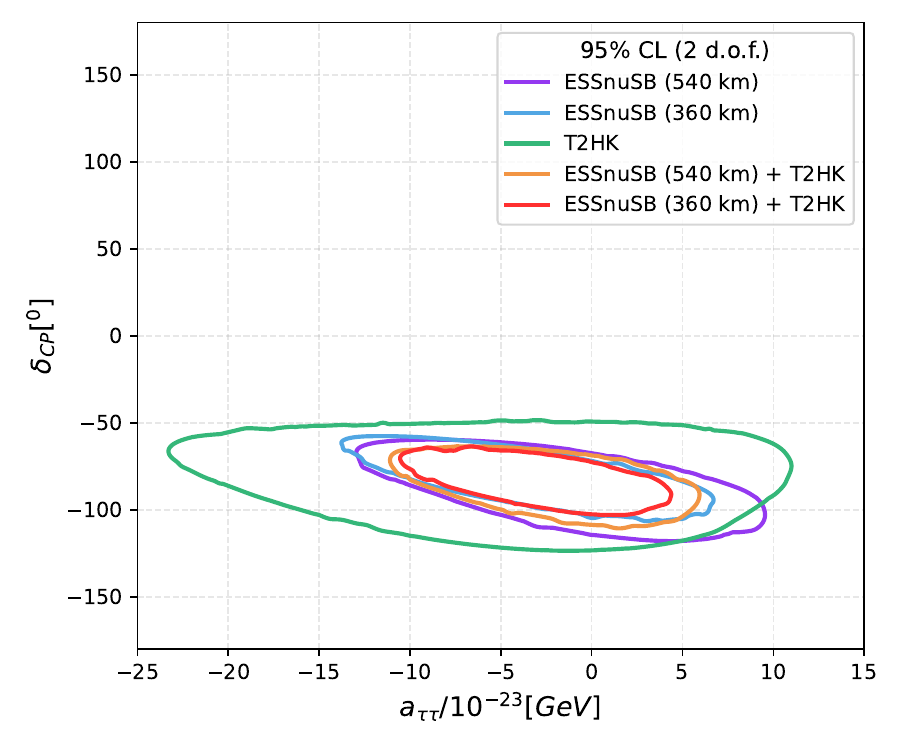}
    \label{fig:subE}
  \end{subfigure}
  \begin{subfigure}[b]{0.48\textwidth}
    \centering
    \includegraphics[width=\textwidth]{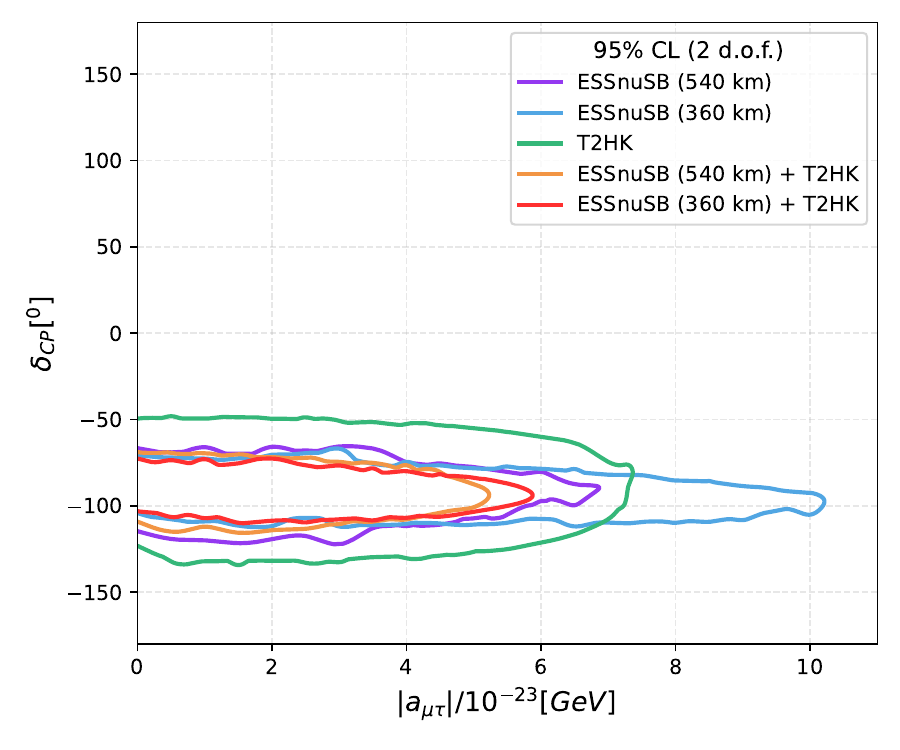}
    \label{fig:subF}
  \end{subfigure}
  \\[-4ex]
\caption{Allowed regions at 95\% C.L. (2 d.o.f.) in the plane of each LIV parameter versus the standard CP phase $\delta_{CP}$. The contours show the sensitivity for ESSnuSB for BL=540 km (purple), ESSnuSB for BL=360 km (cyan), T2HK (green), the combined analysis of ESSnuSB for BL=540 km and T2HK (orange), and the combined analysis of ESSnuSB for BL=360 km and T2HK (red). The true value of all LIV parameters is assumed to be zero, with a true value of $\delta_{CP}$ at maximal CP violation. (true $\delta_{CP} = -90^{\circ}$)}
  
  \label{fig:deltaCP_degeneracy}
\end{figure}

The contours in Figure \ref{fig:theta23_degeneracy} present the 95\% confidence level (C.L.) allowed regions in the two-dimensional parameter space of each LIV parameter versus $\theta_{23}$ for 2 degrees of freedom (d.o.f.). Similarly, contours in Figure \ref{fig:deltaCP_degeneracy} presents the same for $\delta_{CP}$. The left panel presents the results for the diagonal LIV parameters, while the right panel displays the results for the absolute values of the off-diagonal LIV parameters with $\theta_{23}$ and the standard $\delta_{CP}$, respectively. Smaller and more localised allowed regions indicate a stronger ability to constrain both parameters and to resolve potential degeneracies. While generating these plots, for the off-diagonal LIV parameters, we marginalize over the corresponding phase without restriction in its full range, in addition to the standard oscillation parameters as we discussed earlier in Sec \ref{sec:simulation}.

\paragraph{$\mathbf{\theta_{23}}$ vs LIV Parameters:}

An analysis of Figure \ref{fig:theta23_degeneracy} reveals that the longer 540 km ESSnuSB baseline yields a substantially wider allowed parameter space than its 360 km counterpart. We see that for all evaluated correlations between the LIV coefficients and $\theta_{23}$, the 540 km configuration fails to effectively rule out most $\theta_{23}$ values across the 95\% confidence level. The superior sensitivity of the 360 km configuration over the 540 km configuration is primarily driven by statistics. For the 360 km baseline, the beam spectrum allows the experiment to sample the second oscillation maximum via the low-energy tail ($\sim$0.25 GeV) as well as the first oscillation maximum via the high-energy tail ($\sim$0.6 GeV). The 540 km configuration is designed to probe the second oscillation maximum. The neutrino flux at the far detector decreases approximately as ($1/L^2$). The resulting increase in statistical uncertainty outweighs the modest enhancement in oscillation phase obtained from the longer baseline, producing broader allowed regions for the LIV parameters. Consequently, the 360 km baseline provides tighter constraints on both the standard oscillation parameters and the LIV coefficients.

The 360 km setup successfully eliminates the lower octant for parameters $a_{\tau\tau}$, $|a_{e\mu}|$, and $|a_{\mu\tau}|$. A significant challenge emerges for $a_{ee}$ and $|a_{e\tau}|$ as we see that while the true value ($\theta_{23}=48.5^\circ$) is permitted, false minima emerge in the opposite octant, heavily obscuring the genuine octant resolution for the 360 km baseline. For LIV magnitudes around $|a_{e\tau}| \approx (5.5-6)\times 10^{-23}$ GeV and $a_{ee} \approx (5,10)\times 10^{-23}$ GeV, valid $\theta_{23}$ solutions span both octants. This creates a severe degeneracy that prevents an accurate determination of the mixing angle. We observe a parallel issue concerning the correlation of $\theta_{23}$ with $a_{\mu\mu}$, where multiple distinct LIV values can satisfy the same mixing angle, eliminating the possibility of a unique parameter determination. 

T2HK admits multiple distinct LIV values satisfying the same mixing angle for $a_{e\tau}$ and $a_{\mu\mu}$ in the upper octant. However, T2HK cannot truly resolve the parameter degeneracy for $a_{ee}$ and $|a_{e\tau}|$ as valid $\theta_{23}$ solutions persist across both octants for some values of both LIV parameters. T2HK favours the ``true'' higher octant for $a_{\tau\tau}$ for most values of this LIV parameter, while octants are successfully resolved for parameters $|a_{e\mu}|$ and $|a_{\mu\tau}|$. 

We notice that a breakthrough occurs when the high-statistics T2HK configuration is synergised with each of the ESSnuSB baselines. Since the standard oscillation parameters and the LIV parameters behave differently across these distinct energy and baseline regimes, the synergetic combination effectively resolves $\theta_{23}$ degeneracies—eliminating the ``wrong octant" fake solutions of individual ESSnuSB baselines entirely for many LIV parameters. We see that the 540 km ESSnuSB baseline individually allowed wider parameter space compared to the 360 km baseline, failing to resolve parameter degeneracies and ``fake" octant solutions for all LIV parameters, but its synergetic consideration with T2HK significantly improved its performance bringing it capability on par with the synergetic consideration of the 360 km ESSnuSB configuration and T2HK. Most notably, while ESSnuSB 360 km + T2HK combination was not able to resolve the octant degeneracy for $a_{ee}$, the ESSnuSB 540 km + T2HK combination successfully resolved it. For $|a_{e\tau}|$, both ESSnuSB+T2HK combinations substantially reduce the allowed region, but a small lower-octant island remains. Thus, $|a_{e\tau}|$ has the most persistent octant degeneracy in the combined analysis.

At the 95\% confidence level, this joint analytical framework combining ESSnuSB and T2HK constrains the vast majority of the previously degenerate parameter space and ensuring a highly precise isolation of the true variables. The explicitly derived limits for $\theta_{23}$ under these various configurations are summarized in Table \ref{tab:liv_theta23_corr_theta}, calculated assuming a true scenario of standard Lorentz symmetry with $\theta_{23}$ fixed at $48.5^\circ$.

\begin{table}[htbp]
\centering
\renewcommand{\arraystretch}{1.3}
\begin{tabular}{|p{1.2cm}|p{2.4cm}|p{2.4cm}|p{2.2cm}|p{2.5cm}|p{2.5cm}|}
\hline
\textbf{Corr. w/} & \textbf{ESSnuSB (360 km)} & \textbf{ESSnuSB (540 km)} & \textbf{T2HK \ \ \ (295 km) }  &  \textbf{ESSnuSB (360 km) + T2HK (295 km) } & \textbf{ESSnuSB (540 km) + T2HK (295 km) } \\
\textbf{Param} & \textbf{$\theta_{23}$ [deg]} & \textbf{$\theta_{23}$ [deg]} & \textbf{$\theta_{23}$ [deg]} & \textbf{$\theta_{23}$ [deg]} & \textbf{$\theta_{23}$ [deg]} \\
\hline \hline

\multirow{2}{*}{$a_{ee}$}       & $[42.2, 44.3] \cup$ & $[41.5, 50.0]$ & $[42.4, 43.7] \cup$ & $[42.6, 43.4] \cup$ & $[47.6, 48.9]$ \\
                                & $[47.0, 49.6]$      &                & $[47.6, 48.9]$      & $[47.7, 48.8]$      & \\
\hline
\multirow{2}{*}{$a_{\mu\mu}$}   & $[44.3, 50.0]$      & $[42.1, 50.0]$ & $[45.6, 49.1]$      & $[45.9, 49.0]$      & $[46.0, 49.1]$ \\
                                &                     &                &                     &                     & \\
\hline
\multirow{2}{*}{$a_{\tau\tau}$} & $[45.2, 50.0]$      & $[41.7, 50.0]$ & $[44.5, 49.2]$      & $[46.0, 48.9]$      & $[46.3, 49.0]$ \\
                                &                     &                &                     &                     & \\
\hline
\multirow{2}{*}{$|a_{e\mu}|$}   & $[47.2, 49.5]$      & $[41.3, 50.0]$ & $[47.6, 48.9]$      & $[47.8, 48.8]$      & $[47.6, 48.9]$ \\
                                &                     &                &                     &                     & \\
\hline
\multirow{2}{*}{$|a_{e\tau}|$}  & $[42.4, 44.3] \cup$ & $[41.0, 50.0]$ & $[42.2, 43.8] \cup$ & $[42.5, 43.7] \cup$ & $[42.6, 43.7] \cup$ \\
                                & $[47.0, 49.5]$      &                & $[47.6, 48.9]$      & $[47.6, 48.9]$      & $[47.6, 48.9]$ \\
\hline
\multirow{2}{*}{$|a_{\mu\tau}|$}& $[45.3, 49.9]$      & $[41.0, 50.0]$ & $[47.4, 49.0]$      & $[47.6, 48.9]$      & $[47.5, 49.0]$ \\
                                &                     &                &                     &                     & \\
  \hline
\end{tabular}
\renewcommand{\arraystretch}{1.0}
\caption{Projected 95\% C.L. allowed ranges for $\theta_{23}$ in the  presence of various LIV parameters.}
\label{tab:liv_theta23_corr_theta}
\end{table}

\textbf{$\mathbf{\delta_{CP}}$ vs LIV Parameters:} Figure \ref{fig:deltaCP_degeneracy} illustrates the 95\% C.L. (2 d.o.f.) allowed regions for each LIV parameter plotted against $\delta_{CP}$ to evaluate the correlation between the standard CP phase and the new physics LIV parameters. 

We see a less degenerate parameter space as compared to the $\theta_{23}$ analysis for all experiments. Except in the cases involving $|a_{e\tau}|$ and $|a_{\mu\tau}|$, the 540 km ESSnuSB baseline yields a less restrictive parameter space than the 360 km configuration. Across all scenarios, the standard CP phase remains tightly localized near its true maximal CP-violating value of $\delta_{CP} = -90^\circ$ which we considered. This is due to the inherent physical advantage of operating at the second oscillation maximum, where the intrinsic CP asymmetry is significantly amplified. This advantage of ESSnuSB is also illustrated by the fact that T2HK, designed to operate near the first oscillation maximum, admits solutions around $\delta_{CP} = +90^\circ$ in addition to the ``true" value of $\delta_{CP} = -90^\circ$ for the parameter $|a_{e\tau}|$. 

\begin{table}[H]
\centering
\renewcommand{\arraystretch}{1.5}
\begin{tabular}{|p{.8cm}|p{2.5cm}|p{2.5cm}|p{2.6cm}|p{2.5cm}|p{2.5cm}|}
\hline
\textbf{Corr. w/} & \textbf{ESSnuSB (360 km)} & \textbf{ESSnuSB (540 km)} & \textbf{T2HK (295 km)} & \textbf{ESSnuSB (360 km) + T2HK (295 km)} & \textbf{ESSnuSB (540 km) + T2HK (295 km)} \\
 & \textbf{$\delta_{CP}$ [deg]} & \textbf{$\delta_{CP}$ [deg]} & \textbf{$\delta_{CP}$ [deg]} & \textbf{$\delta_{CP}$ [deg]} & \textbf{$\delta_{CP}$ [deg]} \\
\hline \hline
\multirow{2}{*}{$a_{ee}$}        
& $[-120.1, -61.5]$     
& $[-125.3, -60.3]$      
& $[-126.1, -48.3]$       
& $[-109.8, -64.7]$  
& $[-114.9, -64.3]$ \\ 
&                       
&                        
&                         
&  
& \\ 
\hline

\multirow{2}{*}{$a_{\mu\mu}$}    
& $[-112.1, -70.3]$     
& $[-123.3, -66.3]$      
& $[-125.8, -47.8]$       
& $[-108.5, -71.1]$  
& $[-113.7, -67.1]$ \\ 
&                       
&                        
&                         
&  
& \\ 
\hline

\multirow{2}{*}{$a_{\tau\tau}$}  
& $[-108.2, -57.1]$     
& $[-119.3, -59.5]$      
& $[-124.9, -47.9]$       
& $[-104.6, -63.1]$  
& $[-112.2, -63.1]$ \\ 
&                       
&                        
&                         
&  
& \\  
\hline

\multirow{2}{*}{$|a_{e\mu}|$}    
& $[-105.7, -70.3]$     
& $[-120.9, -65.5]$      
& $[-127.8, -44.6]$       
& $[-104.1, -71.1]$  
& $[-111.7, -67.5]$ \\ 
&                       
&                        
&                         
&  
& \\ 
\hline

\multirow{2}{*}{$|a_{e\tau}|$}   
& $[-109.6, -69.6]$     
& $[-116.0, -65.6]$     
& $[-132.4, -28.4] \cup$  
& $[-105.2, -70.4]$  
& $[-110.4, -68.4]$ \\ 
&                       
&                        
& $[24.0, 149.2]$        
&  
& \\
\hline

\multirow{2}{*}{$|a_{\mu\tau}|$} 
& $[-113.3, -65.9]$     
& $[-123.3, -64.3]$      
& $[-135.4, -47.0]$       
& $[-110.9, -71.5]$ 
& $[-116.5, -67.9]$ \\ 
&                       
&                        
&                         
& 
& \\ 
\hline
\end{tabular}
\renewcommand{\arraystretch}{1.0} 
\caption{Projected 95\% C.L. allowed ranges for $\delta_{CP}$ in the presence of various LIV parameters. The combined analysis consistently restricts $\delta_{CP}$ to a tight region around the true maximal CP-violating value ($-90^{\circ}$).}
\label{tab:liv_dcp_corr_dcp_final_cubic}
\end{table} 

However, for $a_{\mu\mu}$, T2HK's exceptional high-statistics data at the first oscillation maximum outperforms the two individual ESSnuSB setups to constrain this LIV parameter. The ESSnuSB (360 km) setup restricts $\delta_{CP}$ to an interval of approximately $42^{\circ}$ width ($[-112.1^{\circ}, -70.3^{\circ}]$), whereas the T2HK interval spans $78^{\circ}$ ($[-125.8^{\circ}, -47.8^{\circ}]$) but T2HK provides strong constraints on the magnitude of the LIV parameters due to its high statistics. This feature was not seen for other parameters. For all other LIV parameters under consideration, T2HK is not marginally better than the two configurations of ESSnuSB. In the capacity to exclude the broader two-dimensional parameter space (spanning both the $\delta_{CP}$ axis and the LIV magnitude axis), T2HK is generally weaker in constraining the admissible values of $\delta_{CP}$ as well as the LIV parameters, with the notable exception of $a_{\mu\mu}$.

Combining T2HK with either of the ESSnuSB's 360 km and 540 km baselines significantly improves the overall capability to constrain the parameter space, with both combinations yielding comparable performance. Interestingly, for the parameter $|a_{e\tau}|$, the stand-alone ESSnuSB configurations exhibit problematic degenerate ``islands" near the true $\delta_{CP}$ value,  which were not resolved even after fusing the complementary strengths of ESSnuSB's second oscillation maximum CP sensitivity with T2HK's first oscillation maximum high-statistics. This can heavily impede accurate parameter extraction for this LIV parameter.

\section{Summary \& Concluding Remarks}\label{sec:conclusion}

In this work, we compared the individual potential of the ESSnuSB long-baseline experiment's two proposed baselines (360 km to Zinkgruvan and 540 km to Garpenberg) to constrain CPT-violating isotropic Lorentz Invariance Violation (LIV) parameters $a_{\alpha\beta}$ with their  synergetic potential when combined with the high-statistics, 295 km baseline of the T2HK experiment. Because non-zero LIV parameters introduce non-standard phase effects, they severely entangle with standard oscillation probabilities. This leads to ``fake solutions" where a new physics signal could be misinterpreted as a standard parameter measurement, or conversely, where standard physics could mask an LIV signal. 

In our investigation, we found that combining T2HK and ESSnuSB provides a highly effective mechanism to resolve these degeneracies, especially for $\theta_{23}$. T2HK, with its beam peaking around the first oscillation maximum for the 295 km baseline, provides high statistics  both in neutrino and antineutrino modes, thanks to its runtime split of 2.5 years in neutrino mode and 7.5 years in antineutrino mode. Meanwhile, ESSnuSB's 540 km baseline, operating near the second oscillation maximum, and the 360 km baseline, which samples both the first and second oscillation maximums, offer a complementary probe with an amplified intrinsic CP asymmetry. The combined analysis removes most wrong-octant solutions, although small residual lower-octant islands remain in selected cases, especially for $|a_{e\tau}|$. For the parameter $a_{ee}$, ESSnuSB 360 km + T2HK combination could not eliminate wrong octant solutions entirely for the lower octant but the 540 km + T2HK combination outperformed the previous combination by breaking the octant degeneracy.

In our recent study \cite{Bora:2025LIV}, we investigated the synergetic performance of ESSnuSB and DUNE. We found that for all LIV parameters, DUNE consistently provides strong constraints on the magnitude of the LIV parameters while ESSnuSB consistently yields narrower allowed intervals for the phase $\delta_{CP}$. From our correlation analysis, we found that the performance of ESSnuSB is superior in preserving the precise value of the $\delta_{CP}$ measurement. This may be attributed to the fact that ESSnuSB operates at the second oscillation maximum, where the intrinsic CP asymmetry is approximately three times larger than at the first oscillation maximum probed by T2HK and DUNE. This amplified CP signal makes the measurement more robust against LIV effects, ensuring that even if LIV parameters are non-zero, they are less likely to mimic a false CP value at ESSnuSB than at T2HK. Table \ref{tab:liv_dcp_corr_dcp_final_cubic} summarises the results from Figure \ref{fig:deltaCP_degeneracy}, showing the allowed ranges of $\delta_{CP}$, when the data are generated assuming exact Lorentz invariance and the true value of $\delta_{CP} = -90^{\circ}$. 

Our results demonstrate that resolving LIV parameter degeneracies does not strictly require the large matter effects of experiments like DUNE for atmospheric mixing angle $\theta_{23}$, and the synergetic performance of T2HK's high statistics at the first oscillation maximum combined with ESSnuSB is sufficient for most LIV parameter cases. However, the combination of ESSnuSB and T2HK is not as effective in constraining the LIV parameter space when compared to the combination of DUNE with ESSnuSB \cite{Bora:2025LIV}. Nevertheless, the synergy between ESSnuSB and T2HK also provides a powerful, matter-independent pathway to constrain Planck-scale LIV physics.

\section*{Acknowledgement}
DD thanks Monojit Ghosh for providing the GLoBES files for the ESSnuSB detector on behalf of the ESSnuSB collaboration.

\bibliographystyle{JHEP}
\bibliography{main}

\end{document}